\documentclass[sigconf]{acmart}

\AtBeginDocument{%
  }

\copyrightyear{2025}
\acmYear{2025}
\setcopyright{acmlicensed}\acmConference[UbiComp Companion '25]{Companion of the the 2025 ACM International Joint Conference on Pervasive and Ubiquitous Computing}{October 12--16, 2025}{Espoo, Finland}
\acmBooktitle{Companion of the the 2025 ACM International Joint Conference on Pervasive and Ubiquitous Computing (UbiComp Companion '25), October 12--16, 2025, Espoo, Finland}
\acmDOI{10.1145/3714394.3750714}
\acmISBN{979-8-4007-1477-1/2025/10}

\usepackage{siunitx}

\begin{document}

\title{Full-body WPT: wireless powering with meandered e-textiles}

\author{Ryo Takahashi}
\authornote{Correspondence to Ryo Takahashi}
\affiliation{%
  \institution{The University of Tokyo}
  \city{Tokyo}
  \country{Japan}
  }
\email{takahashi@akg.t.u-tokyo.ac.jp}
\orcid{0000-0001-5045-341X}

\author{Takashi Sato}
\affiliation{%
  \institution{National Institute of Advanced Industrial Science and Technology (AIST)}
  \city{Saga}
  \country{Japan}
  }
\email{machotakashi-satou@aist.go.jp}
\orcid{0000-0002-8860-0531}

\author{Wakako Yukita}
\affiliation{%
  \institution{The University of Tokyo}
  \city{Tokyo}
  \country{Japan}
  }
\email{yukita@bhe.t.u-tokyo.ac.jp}
\orcid{0009-0003-4043-2318}

\author{Tomoyuki Yokota}
\affiliation{%
  \institution{The University of Tokyo}
  \city{Tokyo}
  \country{Japan}
  }
\email{yokota@ntech.t.u-tokyo.ac.jp}
\orcid{0000-0003-1546-8864}

\author{Takao Someya}
\affiliation{%
  \institution{The University of Tokyo}
  \city{Tokyo}
  \country{Japan}
  }
\email{someya@ee.t.u-tokyo.ac.jp}
\orcid{0000-0003-3051-1138}

\author{Yoshihiro Kawahara}
\affiliation{%
  \institution{The University of Tokyo}
  \city{Tokyo}
  \country{Japan}
  }
\email{kawahara@akg.t.u-tokyo.ac.jp}
\orcid{0000-0002-0310-2577}

\renewcommand{\shortauthors}{R. Takahashi et al.}

\begin{abstract}
We present \textit{Full-body WPT}, wireless power networking around the human body using a meandered textile coil. 
Unlike traditional inductive systems that emit strong fields into the deep tissue inside the body, the meander coil enables localized generation of strong magnetic field constrained to the skin surface, even when scaled to the size of the human body.
Such localized inductive system enhances both safety and efficiency of wireless power around the body.
Furthermore, the use of low-loss conductive yarn achieve energy-efficient and lightweight design. 
We analyze the performance of our design through simulations and experimental prototypes, demonstrating high power transfer efficiency and adaptability to user movement and posture. 
Our system provides a safe and efficient distributed power network using meandered textile coils integrated into wearable materials, highlighting the potential of body-centric wireless power networking as a foundational layer for ubiquitous health monitoring, augmented reality, and human-machine interaction systems.
\end{abstract}

\ccsdesc[500]{Human-centered computing~Interaction devices}
\ccsdesc[500]{Human-centered computing~Ubiquitous and mobile computing}
\ccsdesc[500]{Hardware~Wireless devices}

\keywords{Internet of textiles, meander coil, wireless power transfer, electronic textiles, wearable, electroless plating, full-body}

\begin{teaserfigure}
  \includegraphics[width=\textwidth]{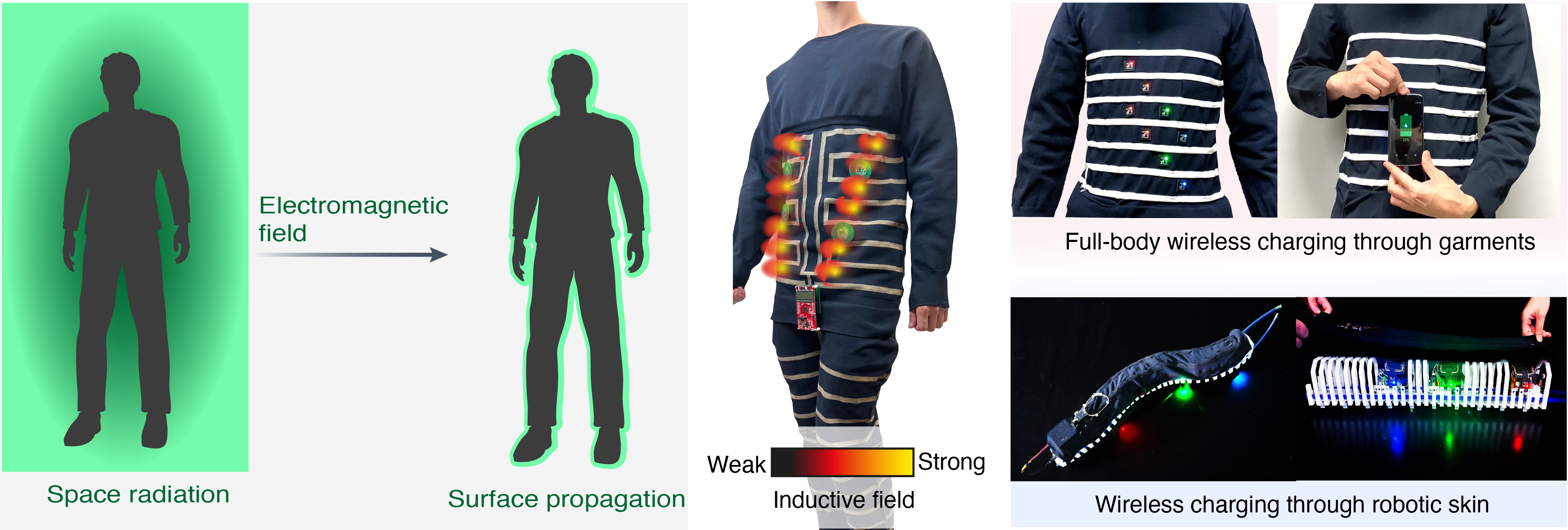}
  \caption{Overview of \textit{Full-body WPT}, wireless powering around the body using meander e-textiles.}
  \Description{}
  \label{fig:teaser}
\end{teaserfigure}

\received{20 February 2007}
\received[revised]{12 March 2009}
\received[accepted]{5 June 2009}

\maketitle

\section{Introduction}

Wearable and implantable devices are playing an increasingly important role in health monitoring, fitness, and human–machine interaction~\cite{negra_wireless_2016}. 
However, powering these devices remains a major challenge. 
Batteries are bulky, need frequent charging, and limit usability for continuous or long-term operation~\cite{lin_digitally-embroidered_2022,tian_wireless_2019}.
Wireless power transfer is a promising alternative, but most existing systems either produce unwanted electromagnetic fields around the body or fail to safely deliver high power~\cite{takahashi_meander_2022}.

We introduce a wireless power network integrated into clothing, using two-dimensional meander coils (see \autoref{fig:teaser}). 
These coils follow a zigzag pattern that reverses direction with each turn, allowing magnetic fields to stay close to the skin even when scaled to full-body size. 
Our system uses either liquid metal~\cite{takahashi_meander_2022,sato_friction_2025} or conductive fiber\cite{takahashi_twin_2022,takahashi_full-body_2025} as the conductor of the textile meander coil, achieving both stretchability and energy efficiency.
This high performance allow for continuous power delivery across the entire body without sacrificing comfort or mobility.
Full-body WPT enables new applications such as full-body sensing suits~\cite{takahashi_twin_2022,takahashi_full-body_2025}, distributed haptic feedback~\cite{takahashi_meander_2022,sato_friction_2025}, and scalable robotics~\cite{kanada_joint-repositionable_2025}.

\section{Related work}

Various approaches have been explored for wirelessly powering on-body electronics, including far-field radio frequency (RF) radiation~\cite{tian_implant--implant_2023,luo_authenticating_2018}, capacitive coupling~\cite{li_body-coupled_2021,varga_enabling_2018}, and near-field magnetic coupling~\cite{lin_digitally-embroidered_2022,takahashi_cuttable_2018,zhu_robust_2024.hajiaghajani_textile-integrated_2021}. 
RF-based systems using antennas can deliver power over longer distances, but they suffer from low efficiency when targeting small, body-mounted devices, and raise safety concerns due to radiation exposure to the dielectric body.
Capacitive coupling used in body-coupled communication (BCC), leverages the human body as a transmission path for electric fields. 
While effective for data transmission, capacitive systems are highly sensitive to grounding conditions and body posture, and typically deliver very limited power due to the safety~\cite{li_body-coupled_2021,takahashi_cuttable_2018}.

In contrast, magnetic coupling via coils offers a more stable and efficient method for on-body power transfer.
Magnetic fields are less affected by the body’s dielectric properties and can be confined to the surface, reducing energy loss and unintended exposure. 
Prior work has explored magnetic resonant coupling and wearable coil arrays~\cite{li_body-coupled_2021,takahashi_cuttable_2018}, but these systems often rely on rigid components or heavy materials like liquid metal, limiting wearability. 
Our work builds on this foundation by using flexible, planar meander coils that maintain magnetic efficiency while enabling lightweight, body-scale integration. 
This allows for continuous power delivery over large areas of the body without sacrificing safety or comfort.

\section{System design}

\begin{figure}[ht!]
  \centering
  \includegraphics[width=1.0\columnwidth]{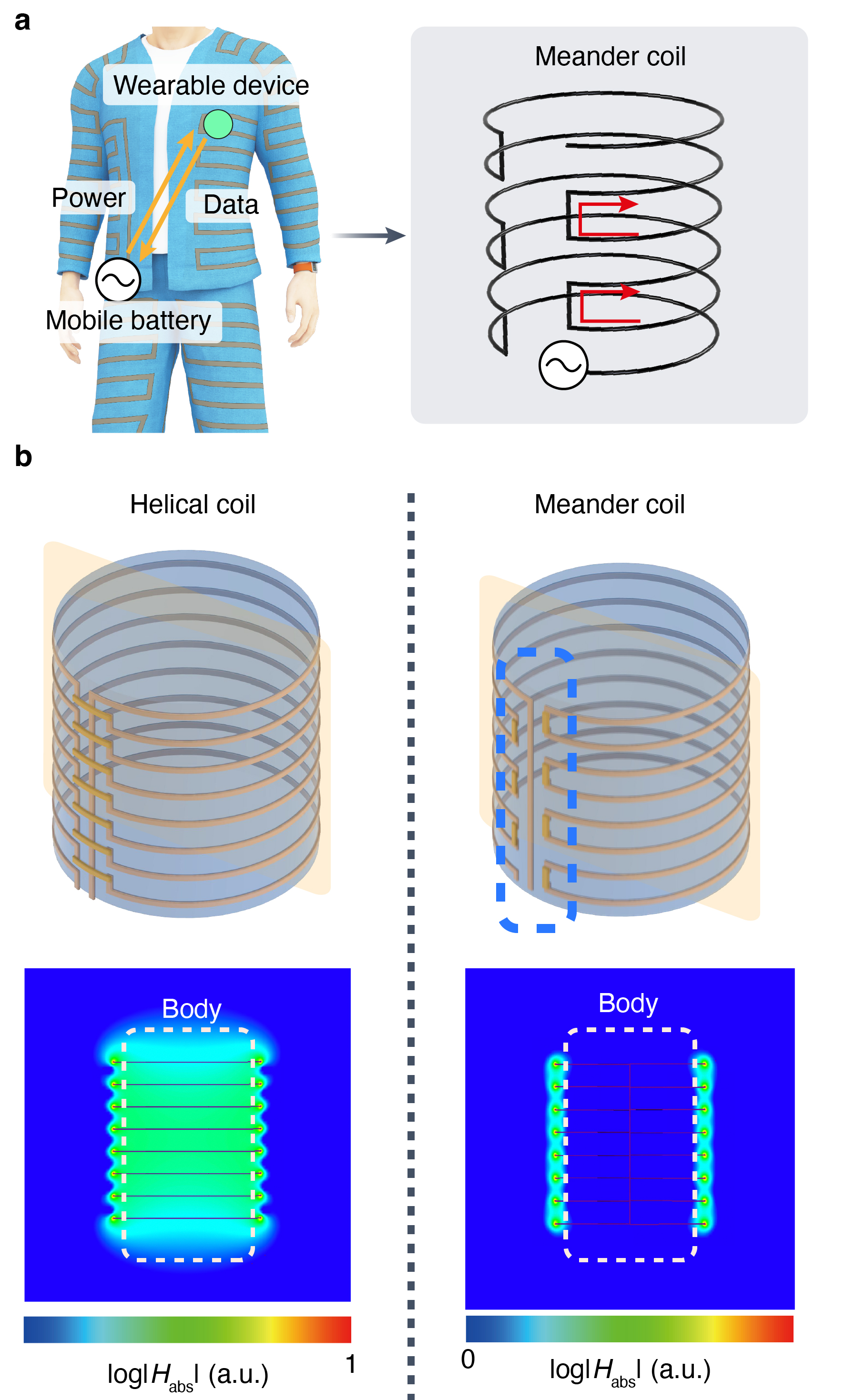}
  \caption{Design of Full-body WPT. (a) Schematic of meander coil. (b) Electromagnetic simulation of meander coil compared to standard helical coil.}
  \label{fig:design}
  \Description{}
\end{figure}

The key component of our wireless power system is a two-dimensional meander coil, designed to generate magnetic fields confined to the skin surface~(see \autoref{fig:design}a). 
The meander coil consists of a serpentine-shaped conductor, in which the current direction alternates with each turn~\cite{takahashi_meander_2022,takahashi_twin_2022}. 
This alternating pattern creates adjacent loops with opposing circulation, resulting in a net magnetic field that is reinforced in the direction parallel to the coil plane and concentrated near the surface.
Unlike traditional circular or spiral coils that generate broad magnetic fields extending into the surrounding space, the meander layout promotes near-field localization, making it suitable for operation close to the body. 
This geometry minimizes stray field leakage and enhances coupling between coils placed on different parts of the body.

The spacing and width of the traces were optimized to maximize magnetic field strength while minimizing resistive loss. 
The resonant frequency of the meander coil is tuned at \SI{13.56}{\MHz}.
An important feature of this design is its scalability.
Multiple meander coils can be distributed over the entire body, forming a continuous or semi-continuous power path that adapts to the user’s posture. 
Because the magnetic field remains close to the surface, this configuration allows power to be routed across the torso, limbs, and joints without significant interference or loss. 
The flat and flexible nature of the coil makes it compatible with integration into textiles, enabling seamless wireless power delivery in wearable garments.

\autoref{fig:design}b shows simulation results for the inductive field generated by two coil types—a body-scale helical coil, and a meander coil, —analyzed using the electromagnetic solver Altair Feko.
The results indicate that the meander coil effectively confines the magnetic field near the skin surface, in contrast to the helical coil, which generates a more dispersed field that extends deeper into the body. 
Therefore, The meander coil further enhances this surface confinement, suggesting a design path for even more localized and efficient on-body power delivery.

\section{Implementation}

\begin{figure}[t!]
  \centering
  \includegraphics[width=0.9\columnwidth]{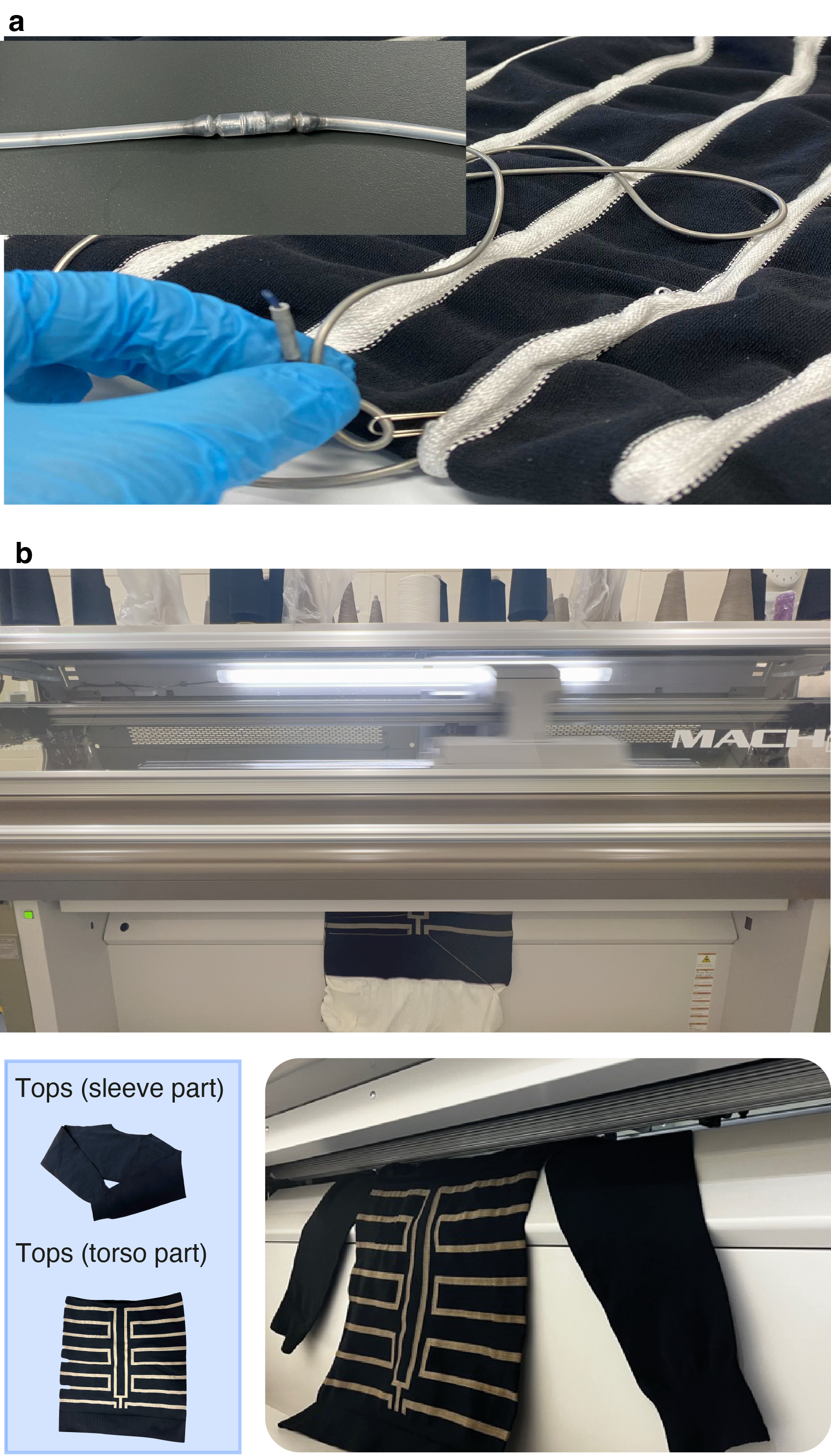}
  \caption{Fabrication of textile meander coil using (a) manual inserting of liquid metal tube or (b) machine knitting of conductive yarn.}
  \label{fig:fab}
  \Description{}
\end{figure}

The meander coil architecture proposed in this work is highly compatible with diverse fabrication techniques. Its zigzag pattern, characterized by repeated directional reversals, is geometrically simple and well-suited for scalable manufacturing.
One approach is to form the coil using liquid metal injected into soft tubing~(see \autoref{fig:fab}a). 
This method enables flexible and stretchable layouts, allowing coils to conform to the shape of the body.
Although the overall system weight increases due to the volume and density of the tubing, this approach remains viable for applications where stretchability is critical.

More importantly, the regular and repetitive structure of the meander coil makes it particularly suitable for machine knitting using conductive yarns~(see \autoref{fig:fab}b). 
Industrial knitting machines can produce long, uniform coil patterns directly into textile structures, enabling seamless integration with garments. 
This method offers high design flexibility, supports mass production, and maintains the softness and wearability required for everyday use.
By leveraging these established textile fabrication methods, the meander coil design can be realized at scale, supporting the deployment of full-body wireless power networks in practical settings.
However, the high resistance of the conductive yarns results in the energy loss of wireless powering even when we use the meander coil structure.
This study implements two types of the meander coils using either liquid metal or conductive yarn, and compares their wireless powering performance against user's motion, i.e., coil deformation.

\section{Power transfer efficiency}

\begin{figure}[ht!]
  \centering
  \includegraphics[width=1.0\columnwidth]{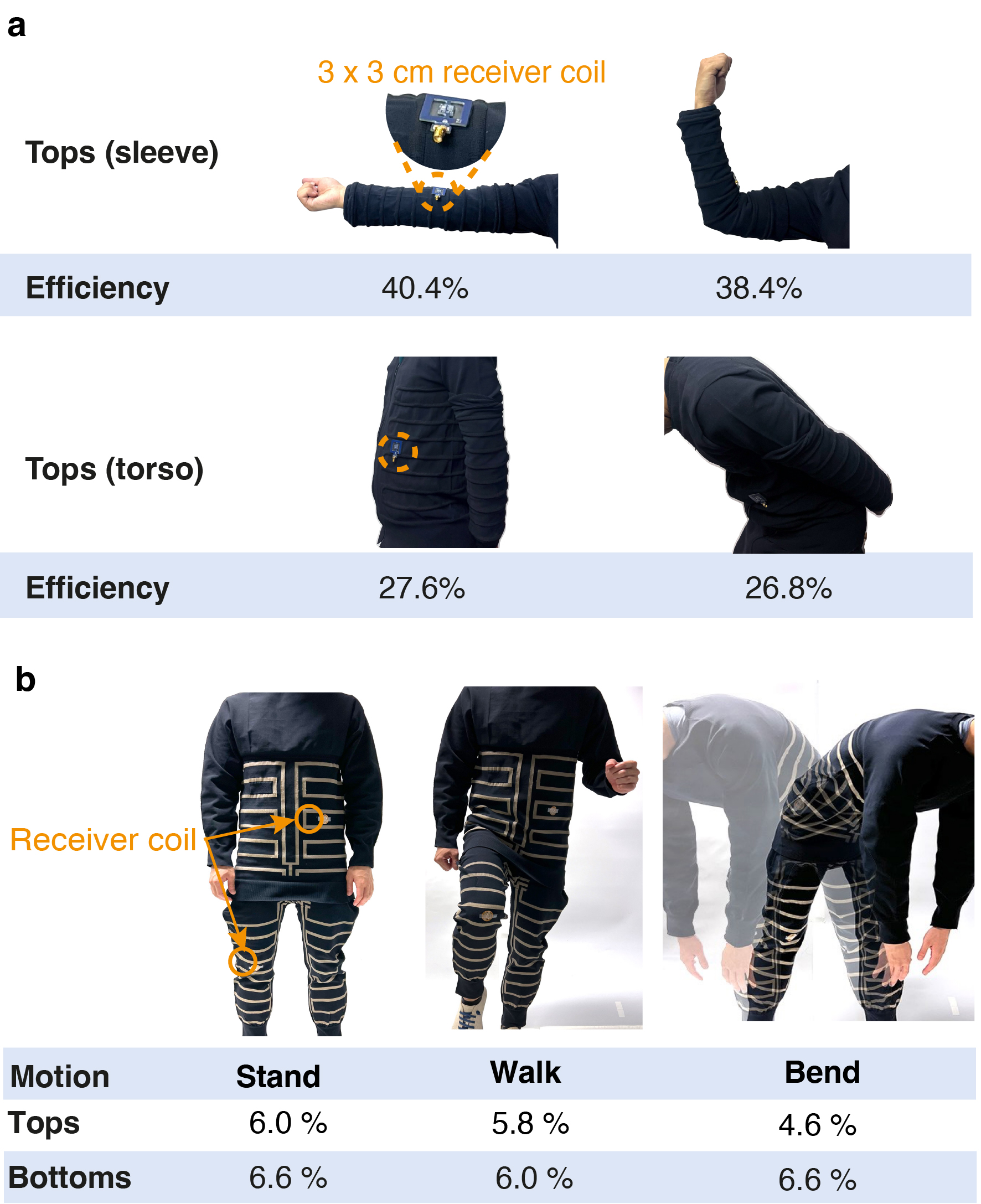}
  \caption{Wireless power performance of (a) liquid-meta-based textile meander coil and (b) conductive-yarn-based meander coil against its coil deformation.}
  \label{fig:power}
  \Description{}
\end{figure}

\begin{figure*}[ht!]
  \centering
  \includegraphics[width=0.9\textwidth]{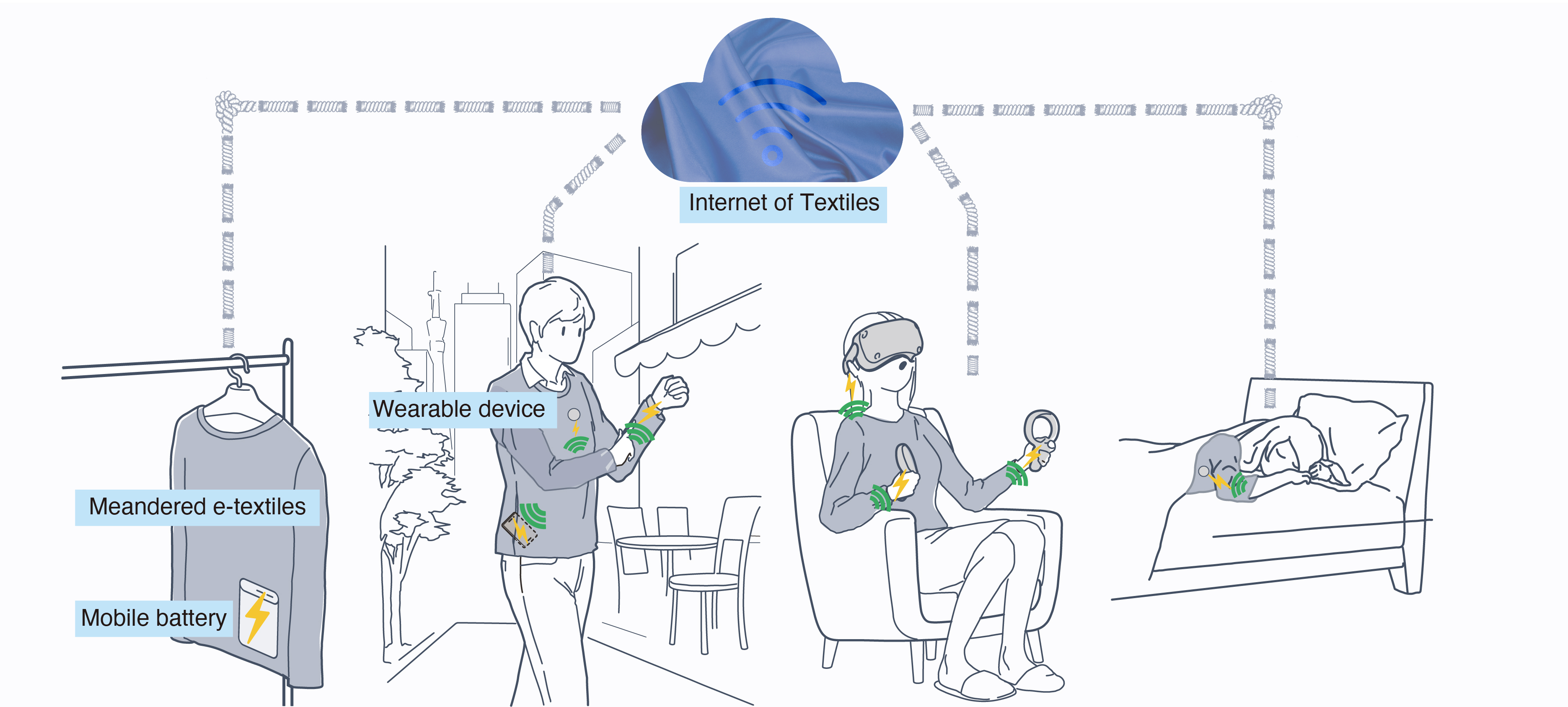}
  \caption{Illustration of Full-body Internet of Textiles using textile-based near-field power and data networks.}
  \label{fig:application}
  \Description{}
\end{figure*}

To evaluate the robustness of the wireless power performance of meandered e-textiles against user's motion, we conducted simple efficiency measurements when the two types of meander coil is deformed. Remarkably, the power transfer efficiency remained almost the same, keeping the relative high efficiency~(see \autoref{fig:power}a). 
Such high efficiency allows watt-class power delivery during the motion, enabling the continuous operation of advanced wearable devices such as wearable bioimager~\cite{yokota_conformable_2020}.
While the conductive yarn lowers the power transfer efficiency~(see \autoref{fig:power}), this still enables the mW-class power delivery, allowing the stable operation of the battery-free NFC sensor tags~\cite{takahashi_full-body_2025}.
Therefore, by selecting the appropriate coil material according to the power requirements of the target device, the meander coil can flexibly support a wide range of wearable applications.

\section{Application examples}

Textiles are ubiquitous in daily life, serving as soft, flexible interfaces between the human body and the environment. Recent advances in stretchable electronics are transforming passive fabrics into electronic textiles (e-textiles) with sensing, actuation, and display capabilities. These developments lay the foundation for the “Internet of Textiles” — large-scale textile networks that collect and share physiological and environmental data to support personal health, behavior change, and smart city infrastructure.

Our wireless power textile platform enables a range of applications in Internet of Textiles. 
As illustrated in \autoref{fig:application}, the system supports seamless power delivery to multiple on-body devices through a network of meandered e-textiles integrated into clothing. 
A small mobile battery can energize the entire network, allowing users to move freely while powering distributed sensors and actuators.
In the home environment, the system can support continuous health monitoring, such as temperature, heart rate, or motion tracking, without requiring battery changes. 
While walking or interacting with devices, wearable sensors can communicate with ambient systems, contributing to smart home integration. During seated activities, such as remote work or gaming, the textile can power haptic feedback actuators in the arms or hands, enabling more immersive interaction. 
At night, the system can continue to deliver power to sleep monitoring devices, maintaining data collection without disturbing the user.

\section{Conclusion}

We presented a body-scale and energy-efficient wireless power network for the human body, using stretchable meander coils integrated into textiles. 
By confining the magnetic field near the skin surface, the system enables safe and continuous power delivery across the entire body, even during motion. 
Our design overcomes key limitations of previous approaches, such as high resistive loss in conductive threads and the excessive weight of liquid metal systems. 
Electromagnetic simulations confirmed strong surface-localized fields and high power transfer efficiency, while real-world applications demonstrated the potential for powering full-body wearable devices.
This work lays the foundation for scalable and comfortable electronic textiles that go beyond sensing to support active functionality. 
Combined with data connectivity~\cite{tian_implant--implant_2023,tian_wireless_2019}, such power networks open new opportunities for the Internet of Textiles—enabling health monitoring, immersive feedback, and responsive environments in everyday life.

\begin{acks}
This work was supported by JST ACT-X JPMJAX21K9, JSPS KAKEN 22K21343, and JST ASPIRE JPMJAP2401. 
\end{acks}

\bibliographystyle{ACM-Reference-Format}
\bibliography{references}

\end{document}